\documentclass[preprint,12pt]{elsarticle}
\usepackage{relsize}
\usepackage{units}
\biboptions{sort&compress}

\usepackage{amssymb}

\usepackage{lineno}

\usepackage[section]{placeins}

\usepackage{color}
\usepackage{graphicx}
\usepackage{subfigure}
\usepackage{amsmath}
\usepackage{xfrac}

\newcommand{\WSRS}{\ensuremath{R^{\mathit{WS/RS}}}}
\newcommand{\pspoint}{\ensuremath{\mathbf{p}}}
\newcommand{\psbarpoint}{\ensuremath{\mathbf{\bar{p}}}}

\newcommand{\intVolume}{\ensuremath{{\Omega}}}
\newcommand{\intVolumeBar}{\ensuremath{\bar{\intVolume}}}
\newcommand{\intVolumeSmall}{\ensuremath{\mathsmaller{\intVolume}}}
\newcommand{\intVolumeBarSmall}{\ensuremath{\mathsmaller{\intVolumeBar}}}
\newcommand{\intOmega}{\ensuremath{ \int\limits_{\intVolume}}}
\newcommand{\intOmegaBar}{\ensuremath{ \int\limits_{\intVolumeBar}}}
\newcommand{\intOmegaInline}{\ensuremath{ \int_{\intVolume}}}

\newcommand{\dphidp}{\ensuremath{\frac{d\Phi}{d\pspoint}}}
\newcommand{\dphidpbar}{\ensuremath{\frac{d\bar{\Phi}}{d\psbarpoint}}}

\newcommand{\deltaP}{\ensuremath{\dphidp\mathrm{d}\pspoint}}
\newcommand{\deltaPbar}{\ensuremath{\dphidpbar\mathrm{d}\psbarpoint}}

\newcommand{\ampA}{\ensuremath{\mathcal{A}}}
\newcommand{\ampB}{\ensuremath{\mathcal{B}}}
\newcommand{\ampAbar}{\ensuremath{\mathcal{\bar{A}}}}
\newcommand{\ampBbar}{\ensuremath{\mathcal{\bar{B}}}}

\newcommand{\ampAp}{\ensuremath{\ampA (\pspoint)}}
\newcommand{\ampAbarp}{\ensuremath{\ampAbar (\psbarpoint)}}
\newcommand{\ampBp}{\ensuremath{\ampB (\pspoint)}}
\newcommand{\ampBbarp}{\ensuremath{\ampBbar (\psbarpoint)}}

\newcommand{\ampApStar}{\ensuremath{\ampA^{*} (\pspoint)}}
\newcommand{\ampAbarpStar}{\ensuremath{\ampAbar^{*} (\psbarpoint)}}
\newcommand{\ampBpStar}{\ensuremath{\ampB^{*} (\pspoint)}}
\newcommand{\ampBbarpStar}{\ensuremath{\ampBbar^{*} (\psbarpoint)}}

\newcommand{\AMag}{\ensuremath{\ampA}}
\newcommand{\BMag}{\ensuremath{\ampB}}
\newcommand{\AbarMag}{\ensuremath{\mathcal{\bar{A}}}}
\newcommand{\BbarMag}{\ensuremath{\mathcal{\bar{B}}}}

\newcommand{\CP}{\ensuremath{\mathit{CP}}}

\newcommand{\CPV}{\ensuremath{\mathrm{CPV}}}

\newcommand{\Imag}{\ensuremath{\mathit{Im}}}
\newcommand{\Real}{\ensuremath{\mathit{Re}}} 

\newcommand{\InterferenceFunction}[1]{\ensuremath{%
 y \Real ( #1 ) + x \Imag ( #1 )%
}}
\newcommand{\InterferenceFunctionStar}[1]{\ensuremath{%
 y \Real ( #1 ) - x \Imag ( #1 )%
}}
\newcommand{\InterferenceFunctionLRB}[1]{\ensuremath{%
 y \Real\!\left( #1 \right) + x \Imag\!\left( #1 \right)%
}}

\newcommand{\Z}{\ensuremath{\mathcal{Z}^f_{\Omega}}}
\newcommand{\Zi}{\ensuremath{\mathcal{Z}^f_{\Omega_i}}}
\newcommand{\Zbar}{\ensuremath{\mathcal{Z}^{\bar{f}}_{\bar{\Omega}}}}
\newcommand{\Zconj}{\ensuremath{\mathcal{Z}^{f*}_{\Omega}}}
\newcommand{\Zbarconj}{\ensuremath{\mathcal{Z}^{\bar{f}*}_{\bar{\Omega}}}}
\newcommand{\ImZ}{\ensuremath{\Imag \Z}}
\newcommand{\ReZ}{\ensuremath{\Real \Z}}

\newcommand{\Zall}{\ensuremath{\mathcal{Z}^f}}

\newcommand{\ImZall}{\ensuremath{\Imag \Zall}}
\newcommand{\ReZall}{\ensuremath{\Real \Zall}}

\newcommand{\CoherenceFactor}{\ensuremath{R_{D}^{f}}}

\newcommand{\AveStrongPhaseDiff}{\ensuremath{\delta_{D}^{f}}}

\newcommand{\ZKpipipi}{\ensuremath{\mathcal{Z}^{K3\pi}}}
\newcommand{\ImZKpipipi}{\ensuremath{\Imag \ZKpipipi}}
\newcommand{\ReZKpipipi}{\ensuremath{\Real \ZKpipipi}}

\newcommand{\rD}{\ensuremath{r_{\mathit{Df}}}}

\newcommand{\ratio}{\rD}        

\newcommand{\phimix}{\ensuremath{\phi_{\mathit{mix}}}}
\newcommand{\complexmix}{\ensuremath{\frac{q}{p}}}
\newcommand{\invcomplexmix}{\ensuremath{\frac{p}{q}}}

\newcommand{\phiCp}{\phimix}

\newcommand{\inlineComplexmix}{\ensuremath{\nicefrac{q}{p}}}

\newcommand{\inlineRmix}{\ensuremath{\left | \inlineComplexmix \right |}}
\newcommand{\inlineRatioCp}{\inlineRmix}

\newcommand{\gamt}{\ensuremath{\Gamma t}}

\newcommand{\gp}{\ensuremath{g_{+}}}
\newcommand{\gm}{\ensuremath{g_{-}}}

\newcommand{\bee}{\ensuremath{b}}

\newcommand{\fp}{\ensuremath{ f_{\pspoint}} }
\newcommand{\fpbar}{\ensuremath{ \bar{f}_{\psbarpoint}} }

\newcommand{\rateDtof}{\ensuremath{ \Gamma\! \left ( \Dtof \right )_{\intVolumeSmall} } }
\newcommand{\rateDtofbar}{\ensuremath{ \Gamma\! \left ( \Dtofbar  \right )_{\intVolumeBarSmall} } }
\newcommand{\rateDbartof}{\ensuremath{ \Gamma\! \left ( \Dbartof  \right )_{\intVolumeSmall} } }
\newcommand{\rateDbartofbar}{\ensuremath{ \Gamma\! \left ( \Dbartofbar  \right )_{\intVolumeBarSmall} } }

\newcommand{\WSRSratio}{\ensuremath{ r_{\intVolumeSmall}(t) }}

\newcommand{\half}{\ensuremath{\frac{1}{2}}}

\newcommand{\bra}[1]{\ensuremath{ \langle #1 |}}
\newcommand{\ket}[1]{\ensuremath{ | #1 \rangle}}

\newcommand{\chisq}{\ensuremath{\chi^2}}

\newcommand{\catchyName}{complex interference parameter}

\newcommand{\zeroCharge}{\ensuremath{\mathrm{0}}}

\newcommand{\te}{\ensuremath{\mbox{}^{th}}}

\newcommand{\cleoc}{\mbox{{C}{L}{E}{O}{-}{c}}}

\newcommand{\gam}{\ensuremath{\gamma}}

\newcommand{\prt}[1]{\ensuremath{\mathit{#1}}}
\newcommand{\particle}[1]{\ensuremath{\mathit{#1}}}

\newcommand{\delKpipipi}{\ensuremath{\delta_D^{K3\pi}}}

\newcommand{\RKpipipi}{\ensuremath{R_D^{K3\pi}}}

\newcommand{\Bpm}{\particle{B^{\pm}}}
\newcommand{\B}{\particle{B^\zeroCharge}}

\newcommand{\D}{\prt{D}}
\newcommand{\Do}{\prt{D^\zeroCharge}}

\newcommand{\Dobar}{\prt{\overline{D}^\zeroCharge}}
\newcommand{\Dzero}{\Do}       
\newcommand{\DzeroBar}{\Dobar} 

\newcommand{\Done}{\prt{D^\zeroCharge_{1}}}
\newcommand{\Dtwo}{\prt{D^\zeroCharge_{2}}}

\newcommand{\Kspipi}{\prt{K_S\pi^+\pi^-}}

\newcommand{\Kpm}{\prt{K^{\pm}}}

\newcommand{\BDK}{\prt{\Bpm \to D \Kpm}}

\newcommand{\DtoKpi}{\ensuremath{\prt{\D \to K^-\pi^+}}}

\newcommand{\KpipipiDCS}{\ensuremath{\prt{K^+\pi^-\pi^+\pi^-}}}

\newcommand{\DtoKspipi}{\ensuremath{\D \to K_S \pi^+\pi^-}}
\newcommand{\DtoKsKK}{\ensuremath{\D \to K_S K^+ K^-}}

\newcommand{\DtoKpipipi}{\ensuremath{\prt{\D \to \KpipipiDCS}}}

\newcommand{\DtoKpipipiDCS}{\ensuremath{\prt{\Do \to \KpipipiDCS}}}

\newcommand{\KpipiZeroDCS}{\ensuremath{\prt{K^+\pi^-\pi^\zeroCharge}}}

\newcommand{\DtoKpipiZeroDCS}{\ensuremath{\prt{\Do \to \KpipiZeroDCS}}}

\newcommand{\Dtof}{\ensuremath{  \Dzero(t) \to f } }
\newcommand{\Dtofbar}{\ensuremath{  \Dzero(t) \to \bar{f} } }
\newcommand{\Dbartof}{\ensuremath{  \DzeroBar(t) \to f } }
\newcommand{\Dbartofbar}{\ensuremath{  \DzeroBar(t) \to \bar{f} } }

\newcommand{\BtoDK}{\ensuremath{  \Bpm \to \D K^{\pm} } }
\newcommand{\BzerotoDKst}{\ensuremath{  \B \to \D K^{*} } }

\newcommand{\un}[2]{\ensuremath{#1\,\mathrm{#2}}}

\newcommand{\eqnref}[1]{Eq.~\ref{#1}}

\newcommand{\eqnsref}[2]{Eqs.~\ref{#1}~-~\ref{#2}}

\newcommand{\tabref}[1]{Tab.~\ref{#1}}

\newcommand{\secref}[1]{Sec.~\ref{#1}}

\newcommand{\secs}{Sections}

\newcommand{\Figref}[1]{Figure~\ref{#1}}
\newcommand{\figref}[1]{Fig.~\ref{#1}}

\numberwithin{equation}{section}

\journal{Phys Lett B}

\newcommand*\patchAmsMathEnvironmentForLineno[1]{%
  \expandafter\let\csname old#1\expandafter\endcsname\csname #1\endcsname
  \expandafter\let\csname oldend#1\expandafter\endcsname\csname end#1\endcsname
  \renewenvironment{#1}%
     {\linenomath\csname old#1\endcsname}%
     {\csname oldend#1\endcsname\endlinenomath}}%
\newcommand*\patchBothAmsMathEnvironmentsForLineno[1]{%
  \patchAmsMathEnvironmentForLineno{#1}%
  \patchAmsMathEnvironmentForLineno{#1*}}%
\AtBeginDocument{%
\patchBothAmsMathEnvironmentsForLineno{equation}%
\patchBothAmsMathEnvironmentsForLineno{align}%
\patchBothAmsMathEnvironmentsForLineno{flalign}%
\patchBothAmsMathEnvironmentsForLineno{alignat}%
\patchBothAmsMathEnvironmentsForLineno{gather}%
\patchBothAmsMathEnvironmentsForLineno{multline}%
}

\begin{document}

\begin{frontmatter}



\title{Charm mixing as input for model-independent determinations of the CKM phase $\gamma$.} 


\author{Samuel Harnew, Jonas Rademacker}

\address{H H Wills Physics Laboratory, Bristol, UK}

\begin{abstract}
The coherence factor and average strong phase difference of \Dzero\
and \DzeroBar\ decay amplitudes to the same final state play an
important role in the precision determination of the CKM parameter
\gam\ using \BtoDK\ and related decay modes.  So far, this important
input from the charm sector could only be obtained from measurements
based on quantum-correlated $\mathit{D\overline{D}}$ pairs produced at
the charm threshold. We propose to constrain these parameters using
charm mixing, using the large charm samples available at the B
factories and LHCb. We demonstrate for the example of \DtoKpipipi\
that a substantial improvement in the precision of the coherence
factor and average strong phase difference can be obtained with this
method, using existing data.

\end{abstract}




\end{frontmatter}


\section{Introduction}
\label{sec:intro}
In this paper we present a new method of constraining the coherence
factor and average strong phase difference between \Do\ and \Dobar\
decay amplitudes to the same multibody final
state~\cite{Atwood:coherenceFactor}, using input from charm
mixing. 

Charm threshold data~\cite{Libby:2010nu,Briere:2009aa,Insler:2012pm,
Lowery:2009id,CLEO:DeltaKpi} provide important input to the
measurement of the charge-parity (\CP) violating phase \gam\ in \BtoDK,
\BzerotoDKst\ and similar decay modes\footnote{\CP-conjugate decays
are implied throughout, unless stated otherwise. \D\ stands for any
superposition of \Do\ and \Dobar.}~\cite{Atwood:coherenceFactor,
BaBar_uses_us:2011up,ModelIndepGammaTheory,LHCb2012DalitzGamma,
LHCb-CONF-2013-006,LHCb2013GammaCombination,LHCb2013ADSObservation},
where the details of the analysis depend considerably on the final
state of the subsequent \D\ decay~\cite{GLW1, GLW2, ADS, DalitzGamma1,
DalitzGamma2, Rademacker:2006zx}.  The importance of charm threshold
data in this context results from the well-defined superposition
states of \Do\ and \Dobar\ accessible with quantum-correlated
\prt{D\overline{D}} pairs.

Charm
mixing~\cite{CDF:Mixing2008,Belle:Mixing2007,BaBar:Mixing2007,BaBar:Mixing2008,BaBar:Mixing2009,LHCb:Mixing,HFAG2013}
also provides well-defined \Do-\Dobar\ superposition states, which can
be used in a similar manner. Previous studies indicate that for \D\
decays to self-conjugate final states, like \DtoKspipi\ and \DtoKsKK,
datasets much larger than those currently available are required to
significantly improve on the existing constraints from the charm
threshold~\cite{ChrisAndGuy2012}. The interference effects due to \D\
mixing in suppressed decay modes such as \DtoKpipipiDCS\ and
\DtoKpipiZeroDCS\ are enhanced compared to self-conjugate decays. We
propose to exploit this feature, and demonstrate that a substantial
reduction of the uncertainty on the coherence factor and average
strong phase difference in \DtoKpipipi\ is possible with existing
data.

This paper is structured as follows: In \secs~\ref{sec:theory} we
review the mixing formalism for multibody \D\ decays, building on and
extending the treatment presented
in~\cite{Malde:2011mk,Bondar:CharmMixingCP}. We present a unified
description of the mixing-induced interference effects in decays to
self conjugate and non-self conjugate final states. In
\secref{sec:measuringCoherenceFactor} we show how \D-mixing can be
used to constrain the coherence factor and strong phase difference.
Using simulated \DtoKpipipi\ decays we demonstrate that a substantial
improvement in the precision of the coherence factor and average
strong phase difference is possible using existing data; here we are
guided by the expected signal yields from LHCb's 2011
and 2012 data taking period.
In \secref{sec:conclusion}, we conclude.

\section{Mixing Formalism}
\label{sec:theory} 
In this section we review the charm mixing formalism for multibody
decays and its relationship to the interference parameters relevant
for the measurement of \gam\ in \BtoDK\ and similar decay
modes~\cite{Malde:2011mk, ChrisAndGuy2012, Bondar:CharmMixingCP}. We
introduce the \catchyName\ \Zall\ that unifies the formalism for
decays to self-conjugate
states~\cite{ModelIndepGammaTheory,Libby:2010nu,Briere:2009aa} and non
self-conjugates
states~\cite{Atwood:coherenceFactor,Insler:2012pm,Lowery:2009id}. \Zall\
is also particularly convenient for parameterising the
constraints on charm interference effects derived from mixing using
suppressed \D\ decay modes, discussed in
\secref{sec:measuringCoherenceFactor}. Finally, in this section,
we identify the important differences in the formalisms conventionally
used for charm mixing measurements on one hand, and \BtoDK\ and
related measurements on the other.

\subsection{Charm mixing}
The mass eigenstates $\ket{\Done}$ and $\ket{\Dtwo}$, with masses
$M_1, M_2$ and widths $\Gamma_1, \Gamma_2$, are related to the flavour
eigenstates $\ket{\Do}$ and $\ket{\Dobar}$ through
\begin{align}
  \ket{ \prt{D_{1}}} & = p \ket{\Dzero} +  q  \ket{\DzeroBar} , \notag \\
  \ket{ \prt{D_{2}}} & = p \ket{\Dzero} -  q  \ket{\DzeroBar} , 
\label{massStates}
\end{align}
where $p$ and $q$ are complex numbers that satisfy $\left|q\right|^2 +
\left|p\right|^2=1$. We also define
\begin{equation}
M \equiv \frac{M_{1}+M_{2}}{2},\;\;\;
\Gamma \equiv \frac{\Gamma_{1}+\Gamma_{2}}{2},\;\;\;
\Delta M \equiv M_{2}-M_{1},\;\;\;
\Delta \Gamma \equiv \Gamma_{2}-\Gamma_{1}
\end{equation}
and the usual dimensionless mixing parameters
\begin{align}
x & \equiv \frac{\Delta M}{\Gamma} , 
  &  
y & \equiv \frac{\Delta \Gamma}{2\Gamma}.
\label{xy}
\end{align}
The deviation of \inlineRmix\ from $1$ is a measure of \CP\
violation (\CPV) in \D-mixing. The phase $\phimix\equiv \arg \left (
  \frac{ q }{ p } \right )$ is a convention-dependent quantity that is
sensitive to \CPV\ in the interference between mixing and decay -
usually, a phase convention is chosen where $\phimix=0$ in the absence
of \CPV.
In practice we will deal with \D\ mesons that have definite flavour
at creation. These evolve over time $t$ to the following superpositions of
\Do\ and \Dobar:
\begin{align}
\ket{\Do(t)}    &= \gp(t)\ket{\Do}     + \frac{q}{p}\gm(t)\ket{\Dobar} ,
\notag \\
\ket{\Dobar(t)} &= \gp(t)\ket{\Dobar}  + \frac{p}{q}\gm(t)\ket{\Do}, \label{d0bartimeDep}
\end{align}
where $|D^{0}(t)\rangle$ refers to a state that was pure \Do\ at time
$t=0$, while $|\bar{D}^{0}(t)\rangle$ refers to a state that was purely
\Dobar\ at $t=0$. The time-dependent functions $\gm(t)$ and $\gp(t)$
are given by
\begin{align}
\gp(t) &= e^{-iMt-\frac{1}{2}\Gamma t} \;\; \cos\left (\frac{1}{2} \Delta Mt - \frac{i}{4} \Delta \Gamma t  \right ) , \notag\\
\gm(t) &= e^{-iMt-\frac{1}{2}\Gamma t}\; i\sin\left (\frac{1}{2} \Delta Mt - \frac{i}{4} \Delta \Gamma t  \right ).
\end{align}
\subsection{The \catchyName\ \Zall}
For the decay amplitudes of a \D\ flavour eigenstate to a particular
final state $f$, or its \CP\ conjugate $\bar{f}$, we use the following
notation:
\begin{align}
& \ampAp    \equiv \bra{ \fp    } \hat{H} \ket{ \Dzero    } ,
& \ampAbarp \equiv \bra{ \fpbar } \hat{H} \ket{ \DzeroBar } ,
\notag\\
& \ampBp    \equiv \bra{ \fp    } \hat{H} \ket{ \DzeroBar } ,
& \ampBbarp \equiv \bra{ \fpbar } \hat{H} \ket{  \Dzero   } .
 \label{amplitudes}
\end{align}
Here \pspoint\ identifies a point in phase space for the multibody
final state $f$, and \psbarpoint\ identifies the corresponding point
for the \CP-conjugate final state, where all final state momenta and
charges are reversed. In practice we will integrate over finite
phase-space volumes. Following~\cite{Atwood:coherenceFactor} we
therefore define\footnote{Throughout this note * is used to denote the
  complex conjugate, whereas $ \bar{ \ } $ is used to denote the \CP\
  conjugate}
\begin{align}
\intOmega \ampAp \ampApStar \deltaP &\equiv \AMag^{2} ,  &
\intOmegaBar \ampAbarp \ampAbarpStar \deltaPbar &\equiv \AbarMag^{2} ,
\notag \\
\intOmega \ampBp \ampBpStar \deltaP &\equiv \BMag^{2} ,  &
\intOmegaBar \ampBbarp \ampBbarpStar \deltaPbar &\equiv \BbarMag^{2}.  \label{integratedAmp}
\end{align}
We use the symbols \dphidp\ and \dphidpbar\ for the density of
states at \pspoint\ and \psbarpoint\ respectively. The integrals
containing \ampAp\ and \ampBp\ run over the phase space volume
$\intVolume$, and the ones containing \ampAbarp\ and \ampBbarp\ run
over the \CP\ conjugate volume $\intVolumeBar$. These volumes can
encompass all of phase space, or any part thereof. The interference
effects are described by the integrals over the cross terms:
\begin{align}
\frac { \intOmega \ampAp \ampBpStar \deltaP}{\AMag \BMag} &\equiv \Z\
,
& \frac { \intOmegaBar \ampAbarp \ampBbarpStar \deltaPbar}{\AbarMag
  \BbarMag} 
&\equiv \Zbar . \label{coherencefactorZ}
\end{align}
This defines the \catchyName\ \Z\ for the final state $f$ over the
phase space region $\Omega$, and \Zbar, its \CP-conjugate. For
integrals over all phase space we use \Zall, omitting the subscript.
The magnitude of \Z\ is between~$0$ and~$1$. The phase of \Z\
represents a weighted average of the phase difference between the two amplitudes over $\intVolume$.
The parameter \Zall\ is directly related to the
coherence factor \CoherenceFactor\ and average phase
difference \AveStrongPhaseDiff\ introduced
in~\cite{Atwood:coherenceFactor},
\begin{equation}
\label{eqn:coherenceFactor}
\Zall  \equiv  \CoherenceFactor e^{-i \AveStrongPhaseDiff}.
\end{equation}
For binned analyses in decays to self-conjugate final states such as
\Kspipi, these interference effects are usually parametrised instead
by $c_i$ and $s_i$. These are weighted averages of the cosine and the
sine of the phase difference between \Do\ and \Dobar\ decay
amplitudes, taken over a phase-space bin $i$ covering the volume
$\Omega_i$. This formalism was originally introduced
in~\cite{ModelIndepGammaTheory}; we follow the definition of $c_i$ and
$s_i$ used in most subsequent articles~\cite{Belle_uses_CLEO_2011,
  LHCb2012DalitzGamma, Libby:2010nu,Briere:2009aa,
  Bondar:CharmMixingCP}. The $c_i$ and $s_i$ parameters are related to
the \catchyName\ through
\begin{equation}
\Zi \equiv c_i + i\, s_i
\end{equation}
We will continue to use \Z\ as it unifies the formalism for
decays to self-conjugate and non-self conjugate states.
In terms of the parameters defined above, the time-dependent decay
rates are given by
\begin{align}
\rateDtof \! = \half
 \Bigg[  \AMag^{2} \left(  \cosh y \gamt  + \cos x \gamt \right ) 
 +  \BMag^{2}  \left ( \cosh y \gamt  - \cos x \gamt \right )  \left|
   \complexmix \right|^{2} 
 \notag \\
   + \ 2 \AMag \BMag  \left[
  \Real\!\left ( \Z\, \complexmix \right ) \sinh (y \gamt) 
 + \Imag\!\left ( \Z\, \complexmix \right ) \sin (x \gamt ) \right]
\Bigg]  
e^{-\gamt}  , \label{rate1}
\end{align}
\begin{align}
\rateDtofbar \! =  \half  
 \Bigg[  \BbarMag^{2} \left(  \cosh y \gamt  + \cos x \gamt \right ) 
+  \AbarMag^{2}  \left ( \cosh y \gamt  - \cos x \gamt \right )
\left| \complexmix \right|^{2} 
 \notag \\
   + \ 2 \AbarMag \BbarMag  \left[
  \Real \left ( \Zbarconj\,  \complexmix \right ) \sinh (y \gamt) 
+ \Imag \left ( \Zbarconj\, \complexmix \right ) \sin (x \gamt ) \right]
\Bigg] 
e^{-\gamt} , \label{rate2}
\end{align}
\begin{align}
\rateDbartofbar \! =  \half  
 \Bigg[  \AbarMag^{2} \left(  \cosh y \gamt  + \cos x \gamt \right ) 
+  \BbarMag^{2}  \left ( \cosh y \gamt  - \cos x \gamt \right )   \left| \invcomplexmix \right|^{2} 
 \notag \\
    + \ 2 \AbarMag \BbarMag  \left[
  \Real \left ( \Zbar\, \invcomplexmix  \right ) \sinh (y \gamt) 
  + \Imag \left ( \Zbar\, \invcomplexmix \right ) \sin (x \gamt )
\right] \Bigg] 
 e^{-\gamt} ,  \label{rate3}
\end{align}
\begin{align}
 \rateDbartof \! =  \half  
 \Bigg[  \BMag^{2} \left(  \cosh y \gamt  + \cos x \gamt \right ) 
 +  \AMag^{2}  \left ( \cosh y \gamt  - \cos x \gamt \right )  \left|
   \invcomplexmix \right|^{2} 
 \notag \\
   + \ 2 \AMag \BMag  \left[
  \Real\! \left ( \Zconj\, \invcomplexmix \right ) \sinh (y \gamt) 
+ \Imag\! \left ( \Zconj\, \invcomplexmix \right ) \sin (x \gamt ) \right]
\Bigg] 
e^{-\gamt} .  \label{rate4}
\end{align}
Assuming that terms of order~3 and higher in the mixing parameters $x$ and
$y$ are negligible leads to the following expressions:
\begin{align}
\rateDtof \simeq 
& {\Bigg [} 
 \AMag^{2} \left ( 1 + \frac{y^{2} - x^{2}}{4} (\gamt)^{2} \right ) + 
\BMag^{2} \left ( \left| \complexmix \right|^{2}  \frac{x^2 + y^2}{4} (\gamt)^{2} \right ) \notag \\
& + \AMag \BMag \left(\InterferenceFunctionLRB{ \Z \complexmix }\right) (\gamt)
 {\Bigg ]}  e^{-\gamt}\label{simpRate1} 
\\
\rateDtofbar \simeq 
& {\Bigg [} 
  \BbarMag^{2}  \left ( 1 + \frac{y^{2} - x^{2}}{4} (\gamt)^{2} \right ) +
\AbarMag^{2}  \left ( \left| \complexmix \right|^{2}  \frac{x^2 + y^2}{4} (\gamt)^{2} \right )  \notag \\
& + \AbarMag \BbarMag \left(\InterferenceFunctionLRB{ \Zbarconj \complexmix }\right) (\gamt)
 {\Bigg ]} e^{-\gamt} \label{simpRate2} 
\\
\rateDbartofbar \simeq 
& {\Bigg [} 
 \AbarMag^{2}  \left ( 1 + \frac{y^{2} - x^{2}}{4} (\gamt)^{2} \right ) +
\BbarMag^{2} \left (\left| \invcomplexmix \right|^{2}  \frac{x^2 + y^2}{4} (\gamt)^{2} \right ) + \notag \\
& \AbarMag \BbarMag \left(\InterferenceFunctionLRB{ \Zbar \invcomplexmix }\right) (\gamt)
 {\Bigg ]}  e^{-\gamt}\label{simpRate3} 
\\
\rateDbartof \simeq  
& {\Bigg [} 
 \BMag^{2} \left ( 1 + \frac{y^{2} - x^{2}}{4} (\gamt)^{2} \right ) +
\AMag^{2} \left (\left| \invcomplexmix \right|^{2} \frac{x^2 + y^2}{4} (\gamt)^{2} \right )  \notag \\
& + \AMag \BMag \left(\InterferenceFunctionLRB{ \Zconj \invcomplexmix }\right) (\gamt)
 {\Bigg ]} e^{-\gamt} .\label{simpRate4} 
\end{align}
For the remainder of this article we assume, for simplicity, that
\CPV\ in charm is negligible, leading to:
$|\Z|=|\Zbar|$, $\ampA=\ampAbar$ and $\ampB=\ampBbar$ (no direct
\CPV); $\inlineRatioCp = 1.0$ (no \CPV\ in mixing); and $\arg( \Z \complexmix ) = \arg ( \Zbar \invcomplexmix )$ (no
    \CPV\ in the interference between mixing and decay).
Following the usual phase convention, we set $\phiCp$ to zero in the
absence of \CPV\ in the interference between mixing and
decay, leading to $\inlineComplexmix=1$ and $\Z = \Zbar $. 
With this, the expressions in \eqnsref{simpRate1}{simpRate4} simplify to
\begin{align}
\rateDtof \simeq  {\Bigg [} 
& \AMag^{2} \left ( 1 + \frac{y^{2} - x^{2}}{4} (\gamt)^{2} \right ) + 
\BMag^{2} \left ( \frac{x^2 + y^2}{4} (\gamt)^{2} \right ) + \notag \\
& \AMag \BMag \left(\InterferenceFunction{ \Z }\right) (\gamt)
 {\Bigg ]}   e^{-\gamt}, \label{noCPsimpRate1} 
\\
\rateDbartof \simeq  {\Bigg [} 
& \BMag^{2} \left ( 1 + \frac{y^{2} - x^{2}}{4} (\gamt)^{2} \right ) +
\AMag^{2} \left ( \frac{x^2 + y^2}{4} (\gamt)^{2} \right ) + \notag \\
& \AMag \BMag \left(\InterferenceFunctionStar{\Z}\right) (\gamt)
 {\Bigg ]} e^{-\gamt} ,  \label{noCPsimpRate2} 
\end{align}
with identical expressions for the \CP-conjugate processes. Since we have
removed all weak phases, $\arg(\Zall)=-\AveStrongPhaseDiff$ now
represents the average \emph{strong} phase difference.
\subsection{Conventions}
There are two different definitions of the \CP\ operator in use. The
Heavy Flavour Averaging Group (HFAG)~\cite{HFAG2013} uses
\begin{equation}
\CP_{\mathrm{HFAG}} \ket{\Do} = -\ket{{\bar{D}}^{0}},
\end{equation}
which is the convention usually adopted for charm analyses.
In the context of extracting \gam\ from \prt{B\to DK} decays, it is usual
practice to follow the ``ADS'' convention~\cite{ADS},
\begin{equation}
\CP_{\mathrm{ADS}} \ket{\Do} = +\ket{{\bar{D}}^{0}}.
\end{equation}
The choice of convention affects several relevant parameters, which
needs to be taken into account when providing charm input to the
measurement of \gam. The choice of convention decides how the mass
eigenstates $\ket{D_{1}}$ and $\ket{D_{2}}$ defined in
\eqnref{massStates} relate to the \CP\ even and odd eigenstates,
$\ket{D_+}$ and $\ket{D_-}$. In the HFAG convention $\ket{D_1} \approx
\ket{D_{-}}$ and $\ket{D_2} \approx \ket{D_{+}}$ (these relations
become exact in the absence of \CPV).  In the ADS convention it is the
other way around.
The mixing variables $x$ and $y$ are defined in terms of (approximate)
\CP\ eigenstates, $x= \frac{M_{+} - M_{-}} {\Gamma}, y=
\frac{\Gamma_{+} - \Gamma_{-}} {\Gamma}$, where the subscripts $+$ and
$-$ label the masses and widths of the predominantly \CP-even and
\CP-odd mass eigenstates, respectively.
The formalism detailed above, with the mixing parameters defined in
\eqnref{xy}, follows the HFAG convention. Changing this to the ADS
convention implies a simultaneous change $x \to -x$ and $y \to -y$.

The choice of convention also affects the \catchyName\ \Z.
To ensure that the same physical \CP\ even or \CP\ odd state
corresponds to the same wave function (up to a phase), the \ket{\Do}
and \ket{\Dobar} wavefunctions between the two conventions must be
related by
\begin{align}
\ket{\Do}_{\mathrm{ADS}}    & = \;\;\, e^{i\xi} \ket{\Do}_{\mathrm{HFAG}} , \notag\\
\ket{\Dobar}_{\mathrm{ADS}} & =  -     e^{i\xi} \ket{\Dobar}_{\mathrm{HFAG}},
\end{align}
where $\xi$ is an arbitrary phase. As
$\Z  \propto
\intOmegaInline
 \bra{ \fp    } \hat{H} \ket{ \Do    } 
 \bra{ \fp    } \hat{H} \ket{ \Dobar }^{\ast}
\,\deltaP,
$
this implies
\begin{align}
\Zall_{\Omega\;\mathrm{ADS}} & = - \Zall_{\Omega\;\mathrm{HFAG}} ,
\end{align}
which is equivalent to
\begin{align}
R_{D\; \mathrm{ADS}}^{f} &= R_{D\; \mathrm{HFAG}}^{f} &
 c_i^{\mathrm{ADS}} &= -c_i^{\mathrm{HFAG}}     
\notag \\
\delta_{D\; \mathrm{ADS}}^{f} &= \delta_{D\; \mathrm{HFAG}}^{f} + \pi &
 s_i^{\mathrm{ADS}} &= - s_i^{\mathrm{HFAG}}.
\end{align}

\section{Constraining the Coherence Factor and strong phase difference with D Mixing}
\label{sec:measuringCoherenceFactor}
\subsection{Overview}
The dependence of \eqnsref{rate1}{rate4} on \Z\ has usually been taken
to imply that external input on \Z\ is required to extract charm
mixing parameters from multibody \D\
decays~\cite{Bondar:CharmMixingCP, Malde:2011mk,
  ChrisAndGuy2012}. Instead, we intend to use existing measurements of
charm mixing
parameters~\cite{CDF:Mixing2008,Belle:Mixing2007,BaBar:Mixing2007,BaBar:Mixing2008,BaBar:Mixing2009,LHCb:Mixing,HFAG2013}
as input, to constrain \Z\ from charm mixing in multibody
decays~\cite{ChrisAndGuy2012}. This in turn provides important input
to the amplitude model-unbiased measurement of
\gam~\cite{ModelIndepGammaTheory,Libby:2010nu,Briere:2009aa,Atwood:coherenceFactor,Insler:2012pm,Lowery:2009id}. So
far, this type of input has only been accessible at the charm
threshold~\cite{Libby:2010nu,Briere:2009aa,Insler:2012pm,
  Lowery:2009id,CLEO:DeltaKpi}.

In \eqnref{noCPsimpRate1}, the term linear in $t$ (the ``interference
term'') is sensitive to \InterferenceFunction{\Z}, while in
\eqnref{noCPsimpRate2} it is sensitive to
\InterferenceFunctionStar{\Z}, so both \ReZ\ and \ImZ\ can be
extracted.
However, previous studies~\cite{ChrisAndGuy2012} indicate that
datasets much larger than those currently available are required to
provide useful constraints on $\Zi$ (or $c_i$ and $s_i$) from mixing
using self-conjugate decays such as \DtoKspipi\ and \DtoKsKK.

We will demonstrate here that significant improvements on \Zall\ can
be achieved with existing data for the case where \Dtof\
is a ``wrong-sign'' (WS) decay. This is a decay where \AMag\ is a doubly
Cabibbo-suppressed (DCS) amplitude, such as \DtoKpipipiDCS, or
\DtoKpipiZeroDCS. \Dbartof\ is the corresponding
``right-sign'' (RS) decay, where \BMag\ is Cabibbo-favoured (CF). In this
case $\AMag \ll \BMag$.
As a result, for typical decay times $t$, the interference term in the
WS rate (\eqnref{noCPsimpRate1}) is of a similar order of magnitude as
the leading term, $\AMag^2$, providing enhanced sensitivity to
\InterferenceFunction{\Z}.
On the other hand, for the RS rate (\eqnref{noCPsimpRate2}), the
constant term, $\BMag^2$, completely dominates the decay rate and
there is effectively no sensitivity to
\InterferenceFunctionStar{\Z}. In practice we will use the RS rate to
normalise the WS rate, as this cancels many experimental
uncertainties.

\subsection{\Z\ from the mixing-induced interference of DCS and CF
  amplitudes}
In this scenario it is useful to define the ratio of the DCS amplitude
($\AMag^{DCS}$) to the CF amplitude ($\BMag^{CF}$):
\begin{align}
\ratio &\equiv \frac{\AMag^{DCS}}{\BMag^{CF}}
\label{eq:defineAmpRatios}
\end{align}
Neglecting terms of order~4 or higher in the small quantities $x, y$
and $\ratio$ results in the following expression for the ratio of WS to RS
decays as a function of the \D\ decay time $t$:
\begin{align}
\WSRSratio
 &= \rD^{2} + \rD
 \left( y \ReZ + x \ImZ\right)
 \gamt +\frac{x^2  + y^2}{4}(\gamt)^{2}.  \label{ratio3}
\end{align}
An analysis of the time-dependent decay rate ratio will, through the
linear term of \eqnref{ratio3}, provide a measurement of
\begin{align}
   \bee &\equiv\left( y \ReZ + x \ImZ\right) \label{eq:defBee}
\end{align}
The factor $\rD$, which also features in the linear term, can be
obtained in the same analysis from the $0\te$ order term of
\eqnref{ratio3}, and $\Gamma$ has been measured very
precisely~\cite{PDG}. Taking the \D\ mixing parameters $x$ and $y$ as
input, we can translate a measurement of \bee\ into constraints in the
$\ReZ-\ImZ$ plane. A given value of \bee\ corresponds to a line of
slope $\nicefrac{y}{x}$ in the $\ReZ-\ImZ$ plane defined by:
\begin{equation}
\label{eq:ImZReZLine}
 \ImZ = - \frac{y}{x} \ReZ + \frac{\bee}{x} \,.
\end{equation}
 \begin{figure}[ht]
    \begin{minipage}[b]{0.48\linewidth}
        \centering
        \includegraphics[width=\linewidth]{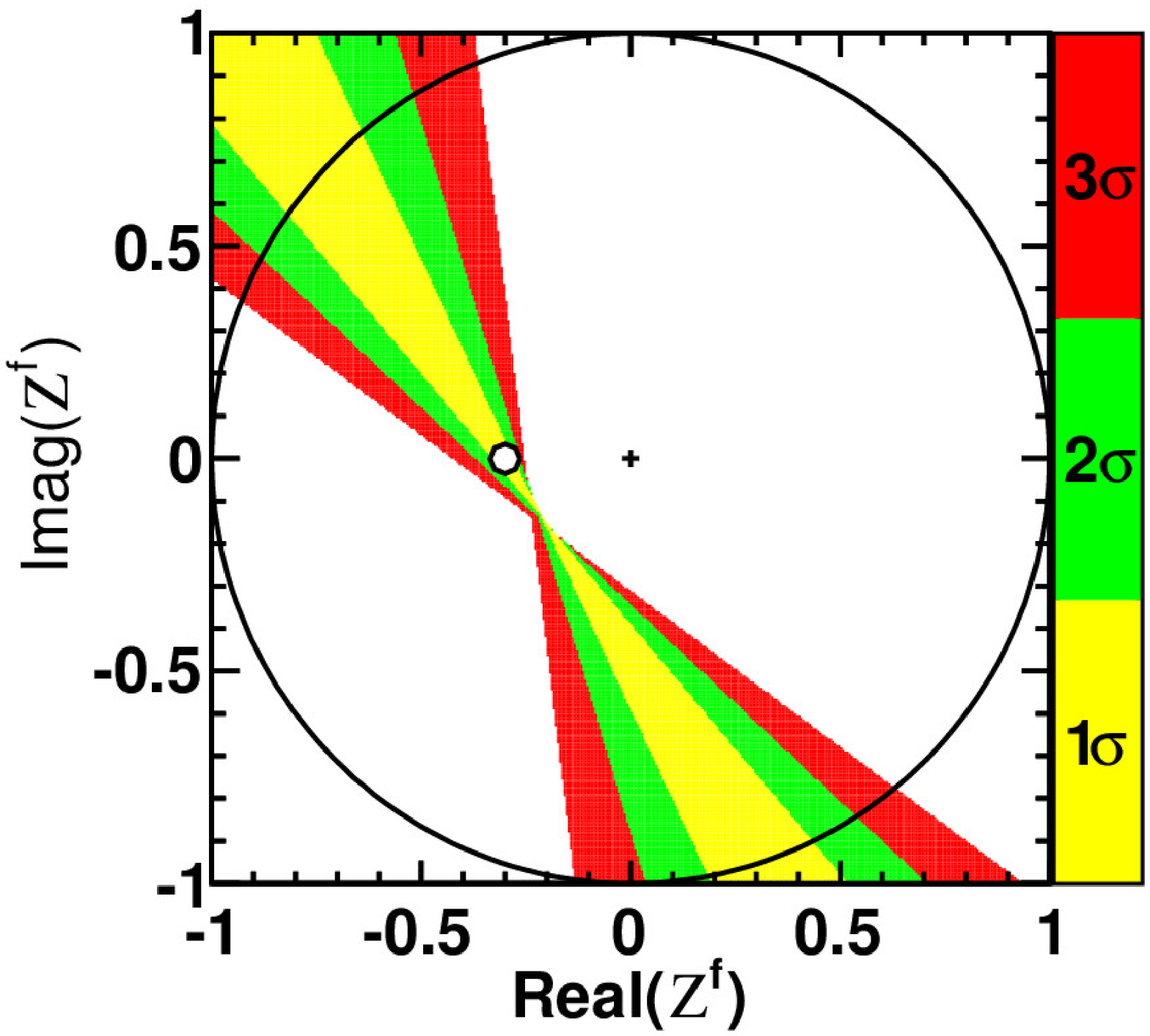}
    \end{minipage}
    \hspace{0.5cm}
    \begin{minipage}[b]{0.48\linewidth}
        \centering
        \includegraphics[width=\linewidth]{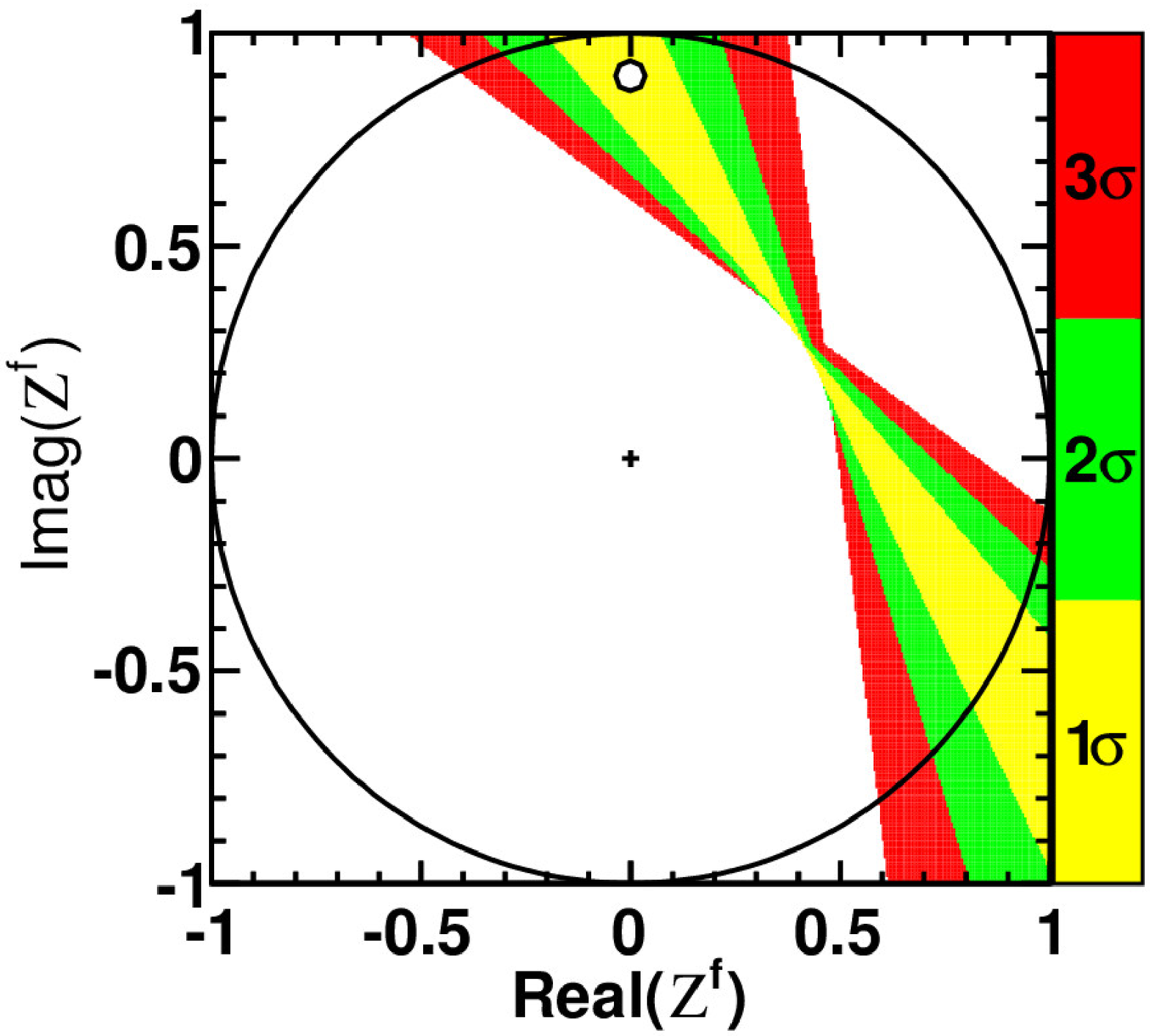}
    \end{minipage}
    \caption{Constraints on \Zall\ for
      $\Zall=-0.3$ (left) and $\Zall=0.9i$ (right), taking
      into account
      current uncertainties on the mixing parameters $x, y$~\cite{PDG}, but ignoring, in
      this illustration, other measurement uncertainties. The white filled
      circle in each plot indicates the central value of \Zall\ used.
      \label{fig:constraints_no_staterror}}
\end{figure}
To show the effect of the current uncertainties in $x$ and $y$ on the
measurement of \Zall\ from \D\ mixing, we consider first the
limiting case of negligible uncertainties on any other parameter, in
particular on \bee\ defined in \eqnref{eq:defBee}. We use the
following values and uncertainties for $x, y$, and their correlation
coefficient $\rho_{x,y}$~\cite{HFAG2013}:
 \begin{equation}
      x = \left(0.63 \pm 0.19 \right)\%
, \;\; y= \left(0.75 \pm 0.12 \right)\%
,\;\;\; \rho_{x,y} = 0.043\,.
\label{eq:xyHFAG}
 \end{equation}
 \Figref{fig:constraints_no_staterror} shows $1, 2$ and $3\sigma$
 confidence limits in the $\ReZall-\ImZall$ plane using these
 inputs for two illustrative example values for the \catchyName, $\Zall =
 -0.3$ and $\Zall = 0.9i$. The $1, 2$ and $3\sigma$ regions are
 calculated using standard techniques based on \chisq\ differences.
\subsection{Sensitivity with existing LHCb datasets}
To estimate the precision on \ZKpipipi\ achievable with current data,
we perform a simulation study based on plausible \DtoKpipipi\ event
yields in LHCb's \un{3}{fb^{-1}} data sample taken in 2011 and
2012. We use the values for the mixing parameters given in
\eqnref{eq:xyHFAG}, and $r_D = 0.058$ based on the WS to RS
branching ratio reported in~\cite{PDG}. We generate simulated events
according to the full expressions for the decay rates given in
\eqnsref{rate1}{rate4}. To take into account the effect of LHCb's
trigger and event selection process, which preferentially selects
decays with long \D\ decay times, we apply a decay-time dependent
efficiency function $\epsilon(t)$ based on that seen
in~\cite{LHCb2011TwoBodyCharmMixing}.  For this feasibility study, we
ignore other detector effects and background contamination. LHCb
results for \DtoKpi\ indicate that backgrounds can be controlled
sufficiently well even for WS decays~\cite{LHCb:Mixing}. Based on the
RS yields reported in~\cite{LHCb2013:Miranda}, and taking into account
that for the WS mode tighter selection criteria might be necessary to
control backgrounds, we estimate about $8$ million RS+WS events in
LHCb's 2011-2012 dataset.  The exact fraction of WS events depends on
the input parameters, in particular on \RKpipipi; typically, $8$
million RS+WS events correspond to about $30,000$ WS events.

To constrain \ZKpipipi\ we perform a \chisq\ fit to the WS/RS ratio in
10 bins of proper decay time. The bins have variable widths, chosen
such that each bin contains a sufficient number of events. Using the
same approximations that led to \eqnref{ratio3}, we obtain for the
expected WS to RS ratio $\WSRS_i$ in bin $i$ that covers the proper
decay time interval $[t^{\mathit{min}}_i, t^{\mathit{max}}_i]$:
\begin{equation}
\WSRS_i
= 
\frac{
\int\limits_{{t^{\mathit{min}}_i}}^{{t^{\mathit{max}}_i}}
\epsilon(t)\, e^{- \Gamma t}\left(
\rD^{2} + \rD
 \left( y \ReZ + x \ImZ\right)
 \gamt +\frac{x^2  + y^2}{4}(\gamt)^{2}\right) \,\mathit{dt}
}{
\int\limits_{{t^{\mathit{min}}_i}}^{{t^{\mathit{max}}_i}}
 \epsilon(t)\, e^{- \Gamma t} \, \mathit{dt}
}
.
\label{eq:binnedRatio}
\end{equation}
The fit parameters are $\ratio$, $b=
\InterferenceFunction{\Z}$, $x$, and $y$, where $x$ and $y$ are
constrained according to \eqnref{eq:xyHFAG}.

\begin{figure}
  \begin{center}
    \subfigure[Number of RS events divided by $e^{-\Gamma t}$,
    representing the shape of $\epsilon(t)$.\label{fig:eff}]{
      \centering
      \parbox[b]{0.45\textwidth}{
      \mbox{\rotatebox{90}{\footnotesize \textsf{\textbf{\mbox{}\hspace{3.5ex}
              Events/$\mathsf{\boldsymbol{e^{-\Gamma t}\,}}$/0.09}}}
      \includegraphics[height=0.62\linewidth]{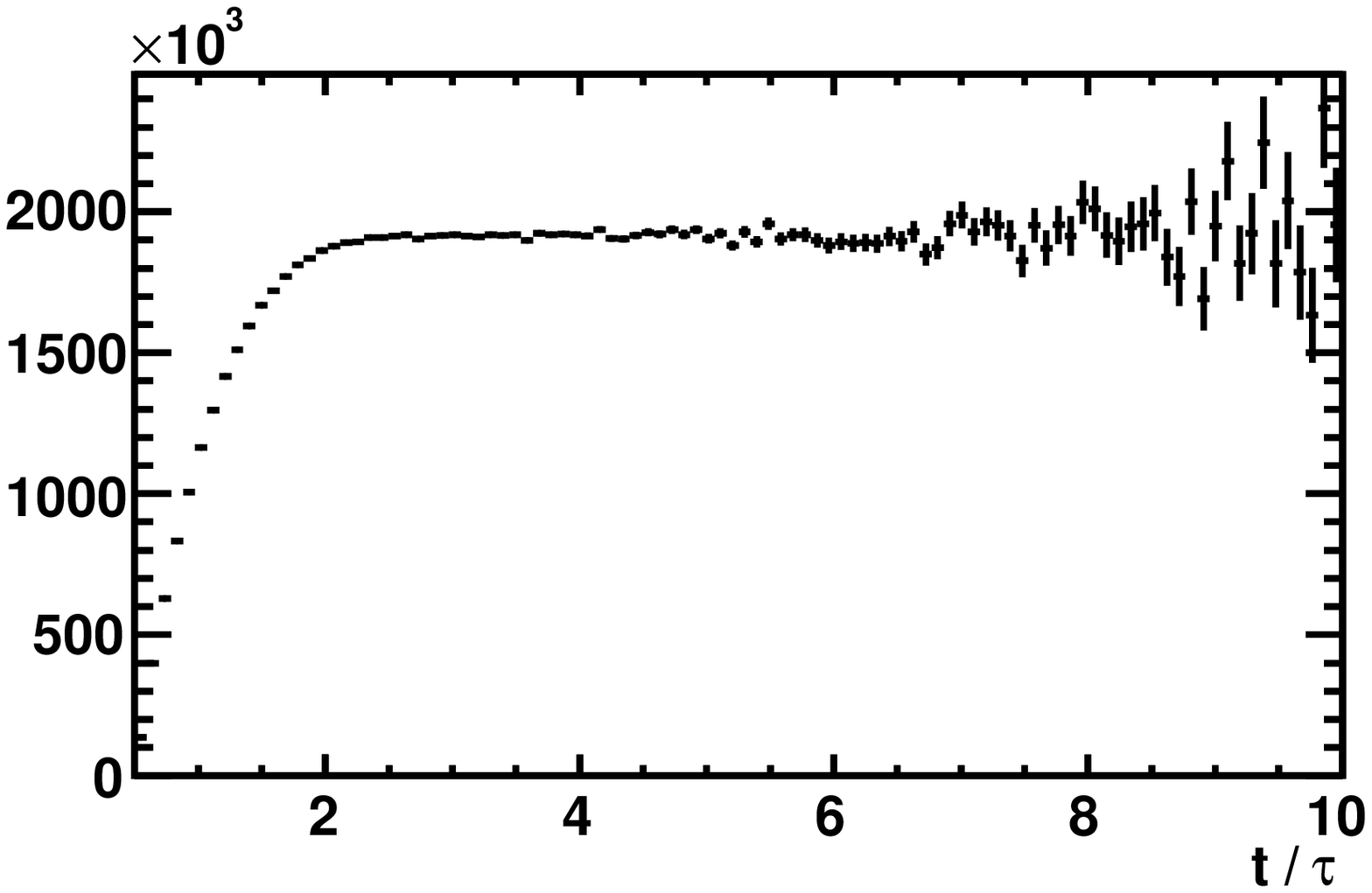}}}
    }
    \hfill
    \subfigure[Ratio of WS to RS events (crosses with error bars) and
    the fit (line).\label{fig:fit}]{
      \centering
      \parbox[b]{0.45\textwidth}{
        \mbox{}\hspace{-0.6em}
      \mbox{\mbox{}\rotatebox{270}{\includegraphics[width=0.64\linewidth]{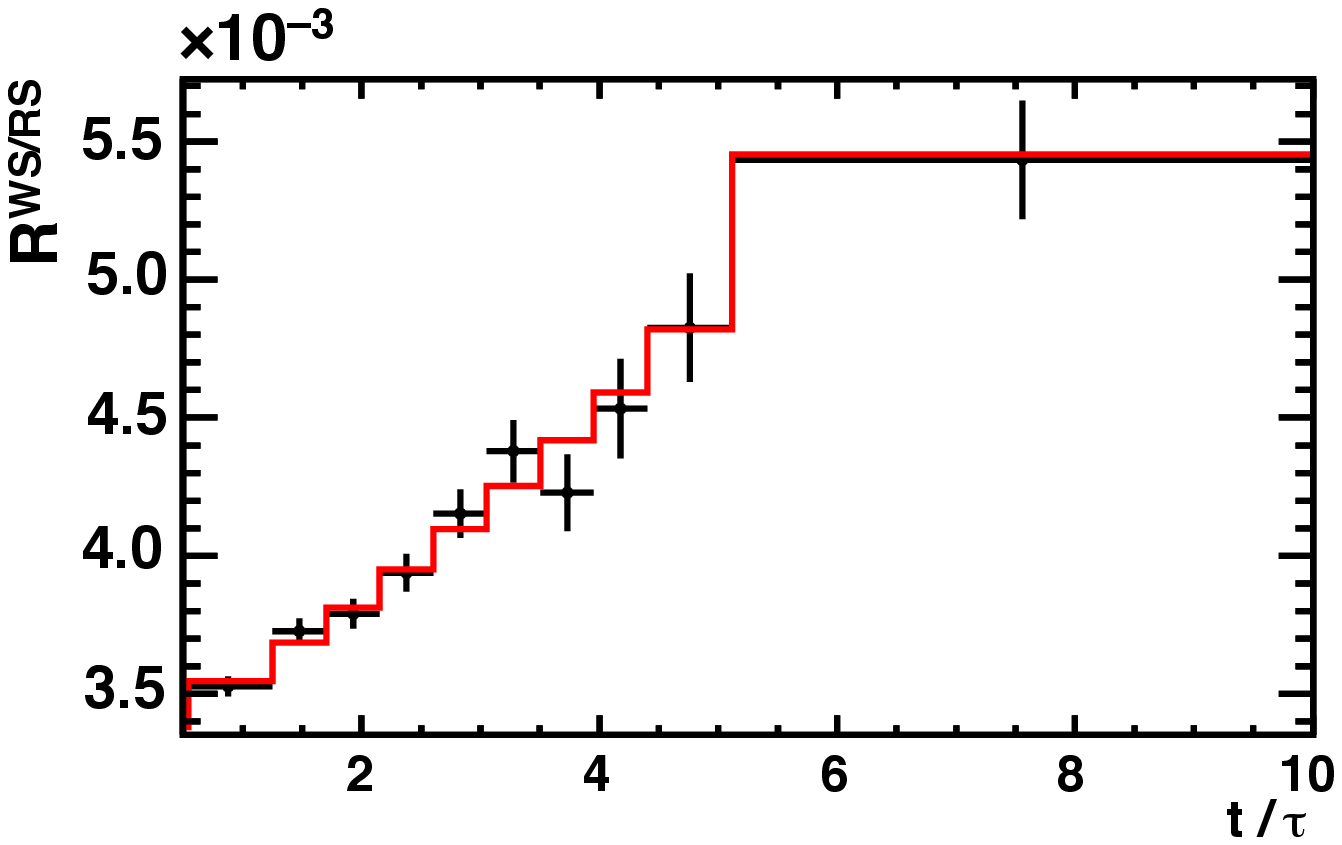}}}
      \vspace{-3.3ex}\\
    }}
    \caption{Simulated data and
      fit in bins of proper decay time, expressed in units of
      $\tau=\nicefrac{1}{\Gamma}$.  The discontinuous shape of the
      line representing the fit in \figref{fig:fit} reflects the way the expected WS/RS
      ratio is calculated for each bin, described in the text.
      \label{fig:effAndfit}}
  \end{center}
\end{figure}
In a real experiment, the time-dependent efficiency $\epsilon(t)$
would not necessarily be known a priory, but it is reasonable to
assume that $\epsilon(t)$ would be the same for WS and RS decays. We
therefore extract its shape from the (simulated) data by dividing the
RS decay time distribution (histogrammed in 100 bins) by $e^{- \Gamma
  t}$, as shown in \figref{fig:eff}; the overall normalisation
cancels when using $\epsilon(t)$ in \eqnref{eq:binnedRatio}.

A pull study based on generating and fitting $200$ simulated data
samples, each containing 8 million RS+WS events, shows no evidence of
fit biases, and confirms the correct coverage of the confidence
intervals obtained from the fit \chisq.

\begin{figure}[ht]
    \begin{minipage}[b]{0.48\linewidth}
        \centering
        \includegraphics[width=\linewidth]{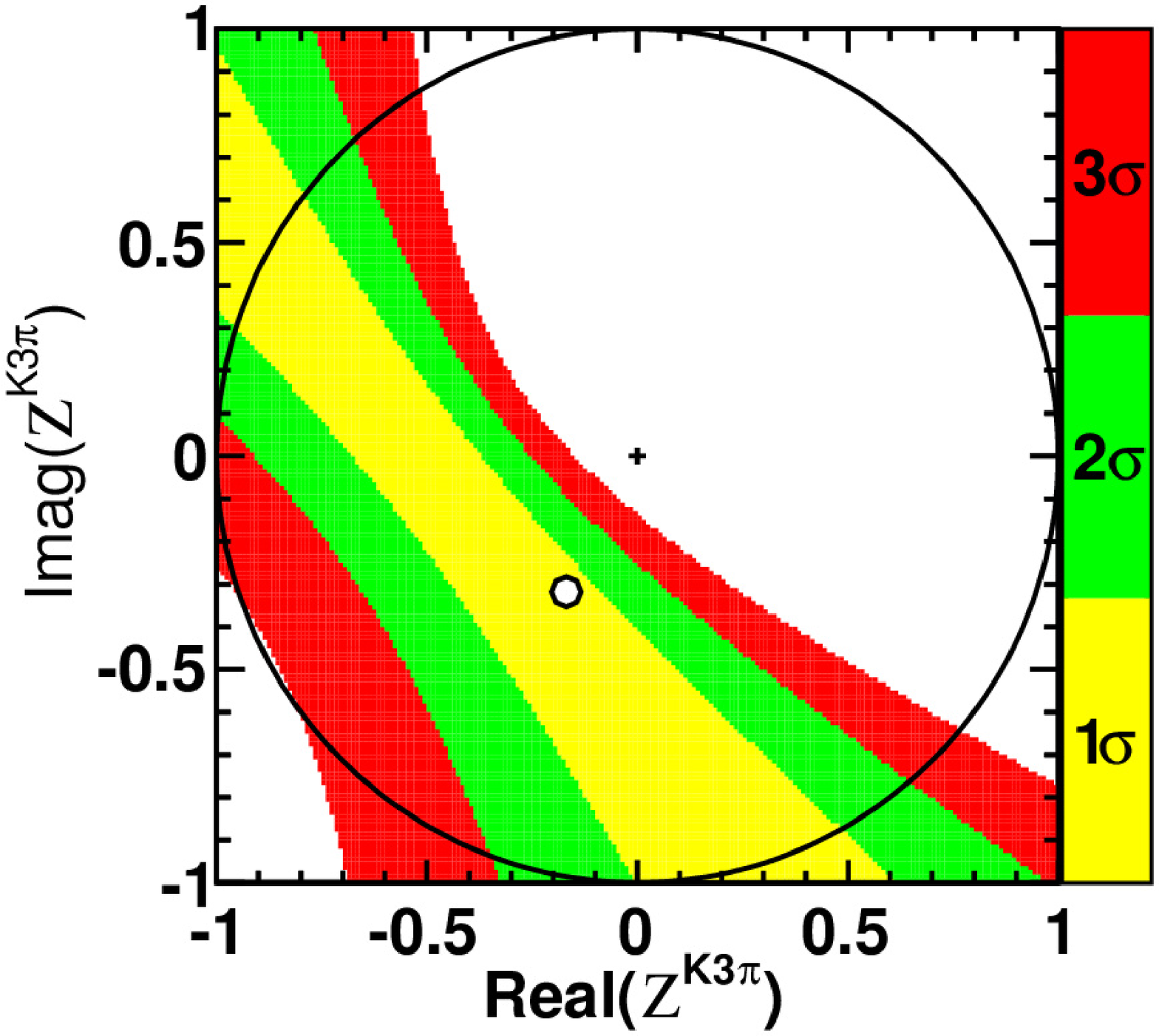}
    \end{minipage}
    \hspace{0.5cm}
    \begin{minipage}[b]{0.48\linewidth}
        \centering
        \includegraphics[width=\linewidth]{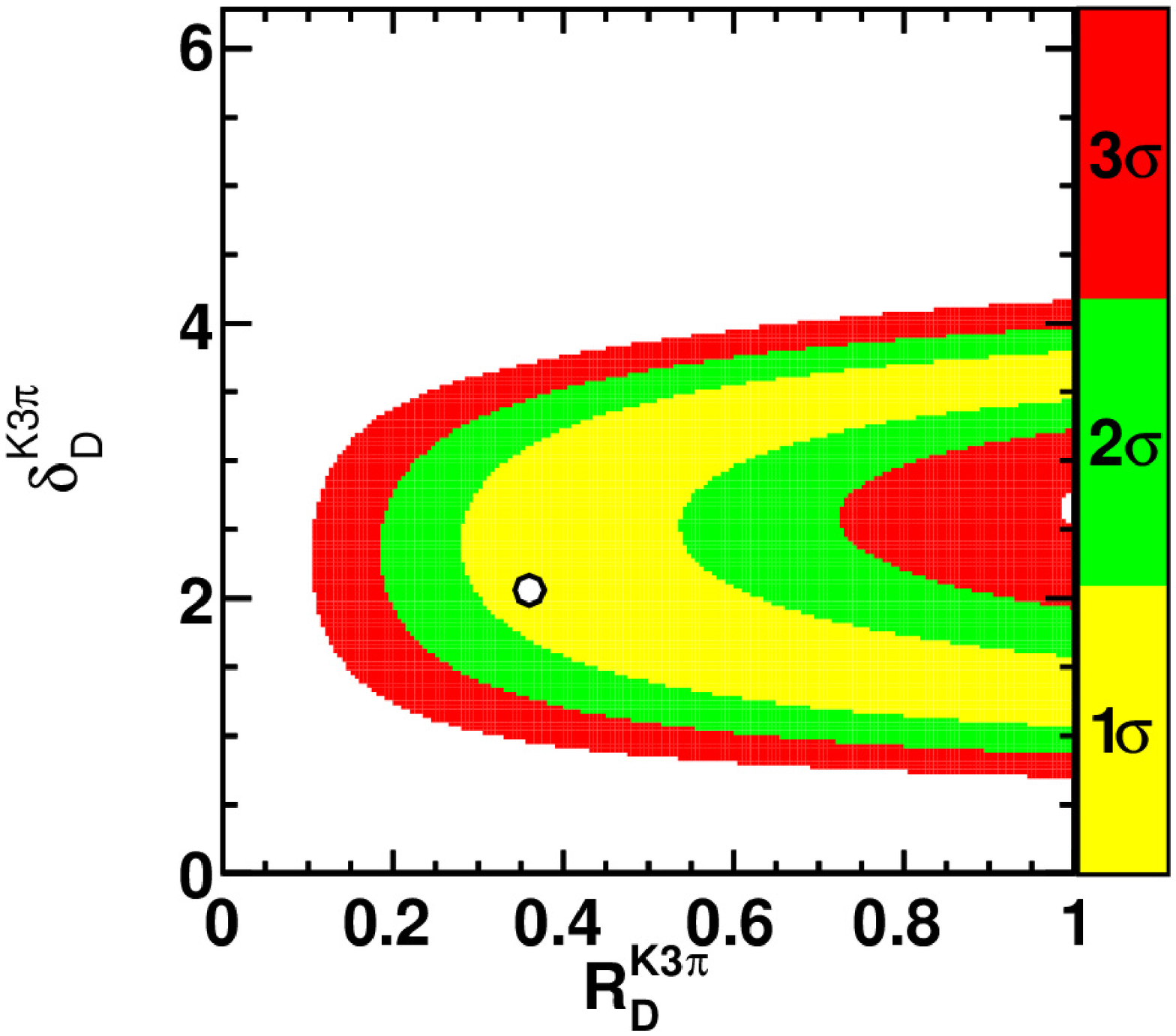}
    \end{minipage}
    \caption{Constraints on \ZKpipipi\ for $8M$ RS and $30k$ WS
      simulated events generated with \cleoc's central value for the
      \catchyName, 
      $\ZKpipipi = -0.133 - 0.301 = 0.33 e^{- 1.99 i}$~\cite{Lowery:2009id}. 
      The constraints are shown both in
      Cartesian (left) and polar coordinates (right). The white filled
      circle indicates the values used to generate the events.}
\label{fig:LHCbconstraints}
\end{figure}

An example of such a fit is shown in \figref{fig:fit}. The $8M$ events have
been generated using \cleoc's central value $\ZKpipipi = {-0.133}
{-0.301 i}$~\cite{Lowery:2009id} and include $30.5k$ WS events.
\begin{figure}[ht]
  \begin{minipage}[b]{0.48\linewidth}
    \centering
    \includegraphics[width=\linewidth]{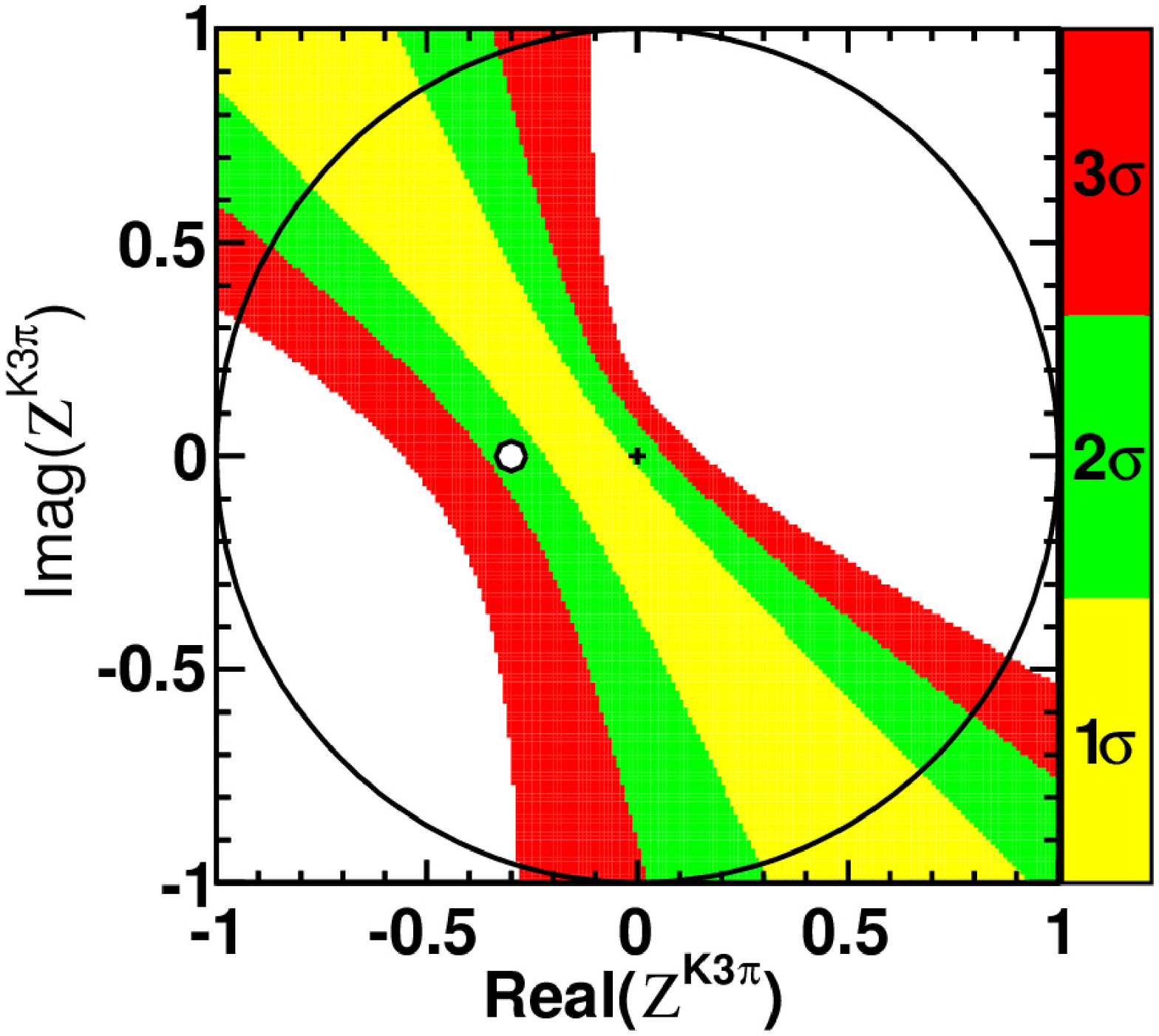}
  \end{minipage}
  \hspace{0.5cm}
  \begin{minipage}[b]{0.48\linewidth}
    \centering
    \includegraphics[width=\linewidth]{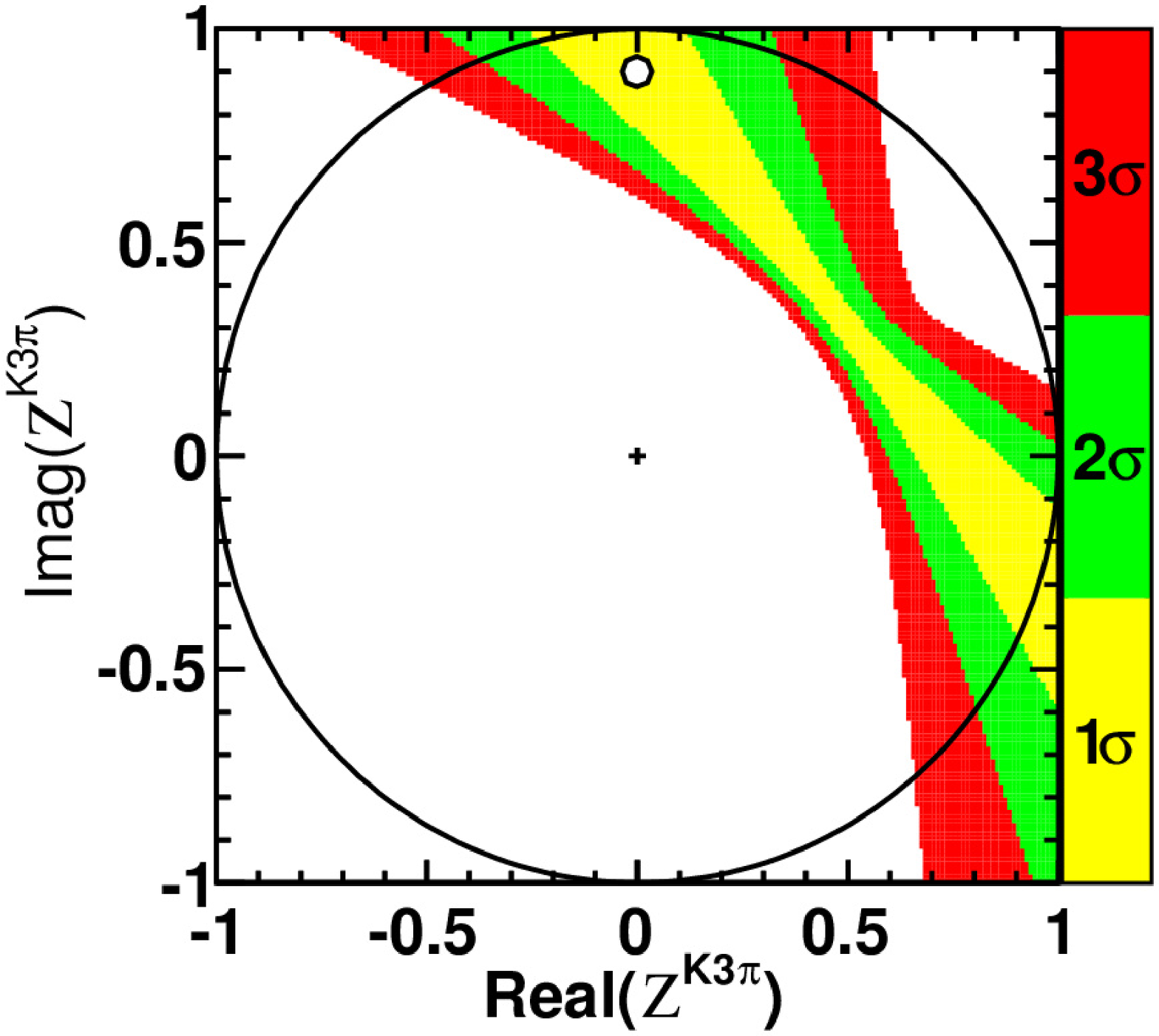}
  \end{minipage}
  \caption{Examples for constraints on \ZKpipipi\ obtained from
    8~million simulated events, generated with $\ZKpipipi=-0.3$ (left)
    and $\ZKpipipi=0.9i$ (right), with current uncertainties on $x,
    y$. The white filled circle indicates the value of
    \ZKpipipi\ used to generate the events.
    \label{fig:constraints}}
\end{figure}
\Figref{fig:constraints} shows $1, 2$ and $3\sigma$ confidence regions
based on 8~million simulated events that have been generated with the
illustrative values $\ZKpipipi=-0.3$ and $\ZKpipipi = 0.9i$ used also
to obtain \figref{fig:constraints_no_staterror}.
\Figref{fig:LHCbconstraints} shows the constraints for events
generated using the \cleoc\ central value for \ZKpipipi, in both polar
coordinates (i.e. the coherence factor $\RKpipipi =
\left|\ZKpipipi\right|$ and strong phase difference
$\delKpipipi=-\arg\left(\ZKpipipi\right)$) and cartesian coordinates
(\ReZKpipipi\ and \ImZKpipipi).

\begin{figure}
    \begin{minipage}[b]{0.48\linewidth}
        \centering
        \cleoc~\cite{Lowery:2009id}
        \includegraphics[width=\linewidth]{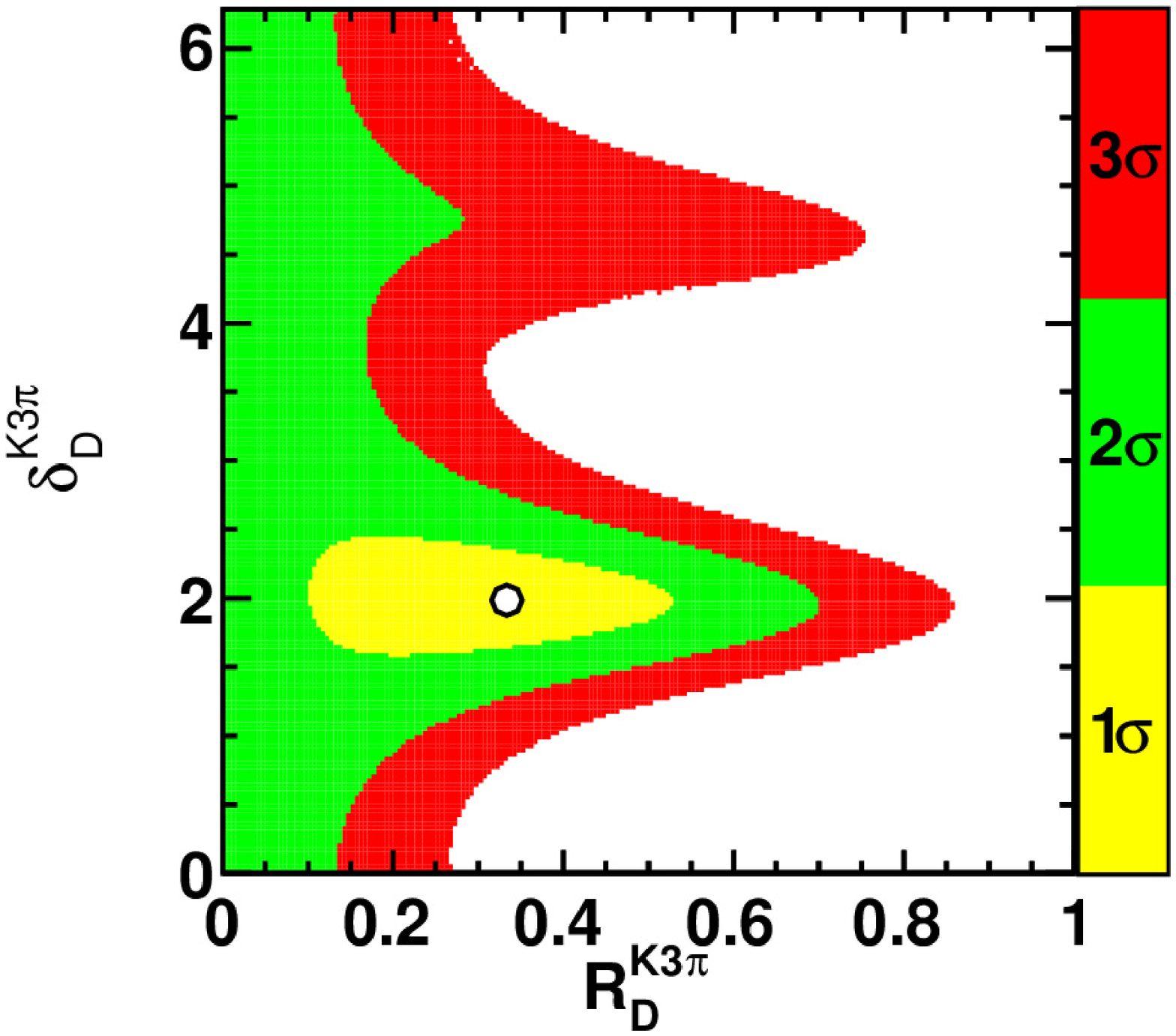}
    \end{minipage}
    \hspace{0.5cm}
    \begin{minipage}[b]{0.48\linewidth}
        \centering
        \cleoc~\cite{Lowery:2009id} combined with simulated input from
        charm mixing
        \includegraphics[width=\linewidth]{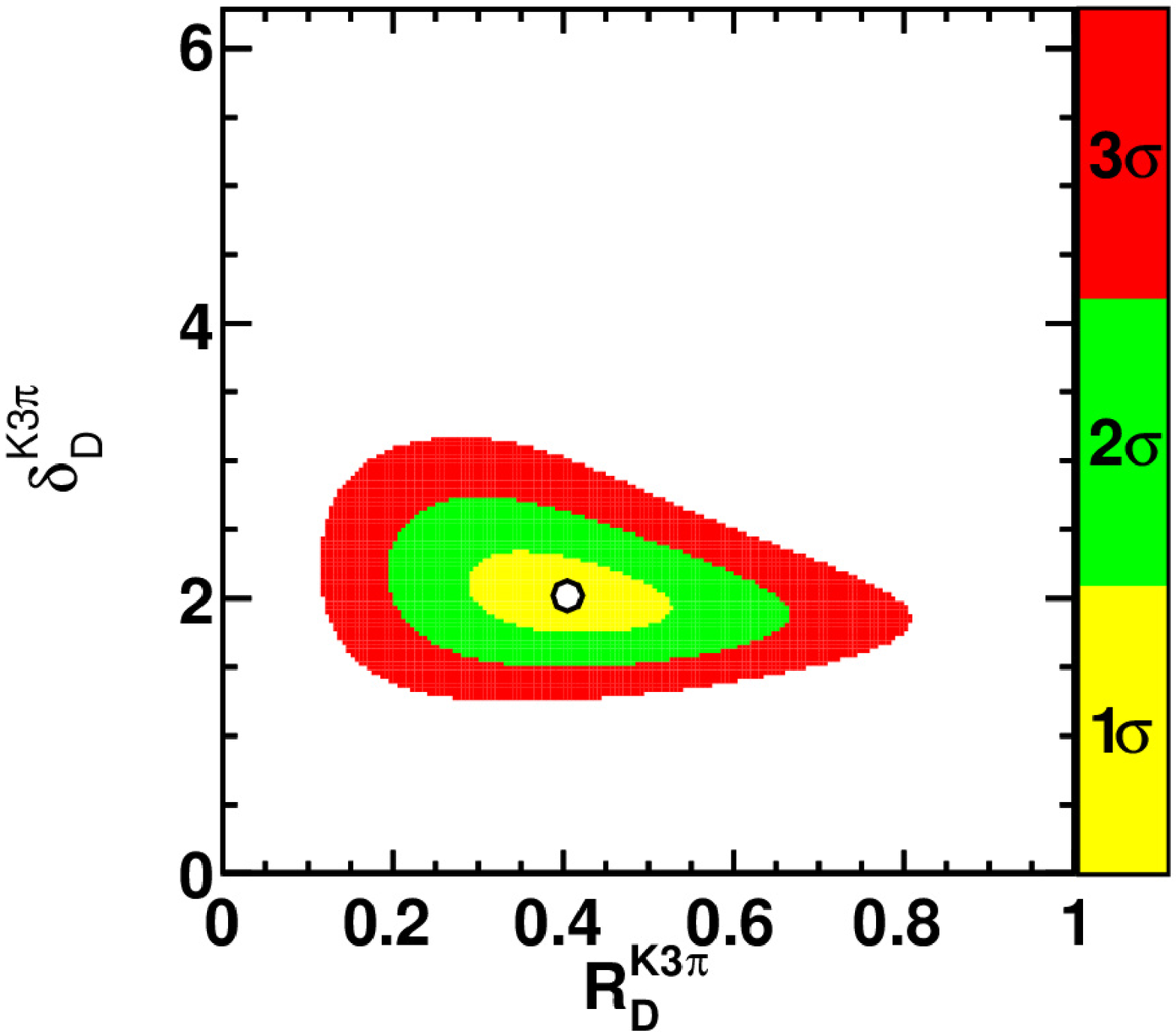}
    \end{minipage}
    \begin{minipage}[b]{0.48\linewidth}
        \centering
        \includegraphics[width=\linewidth]{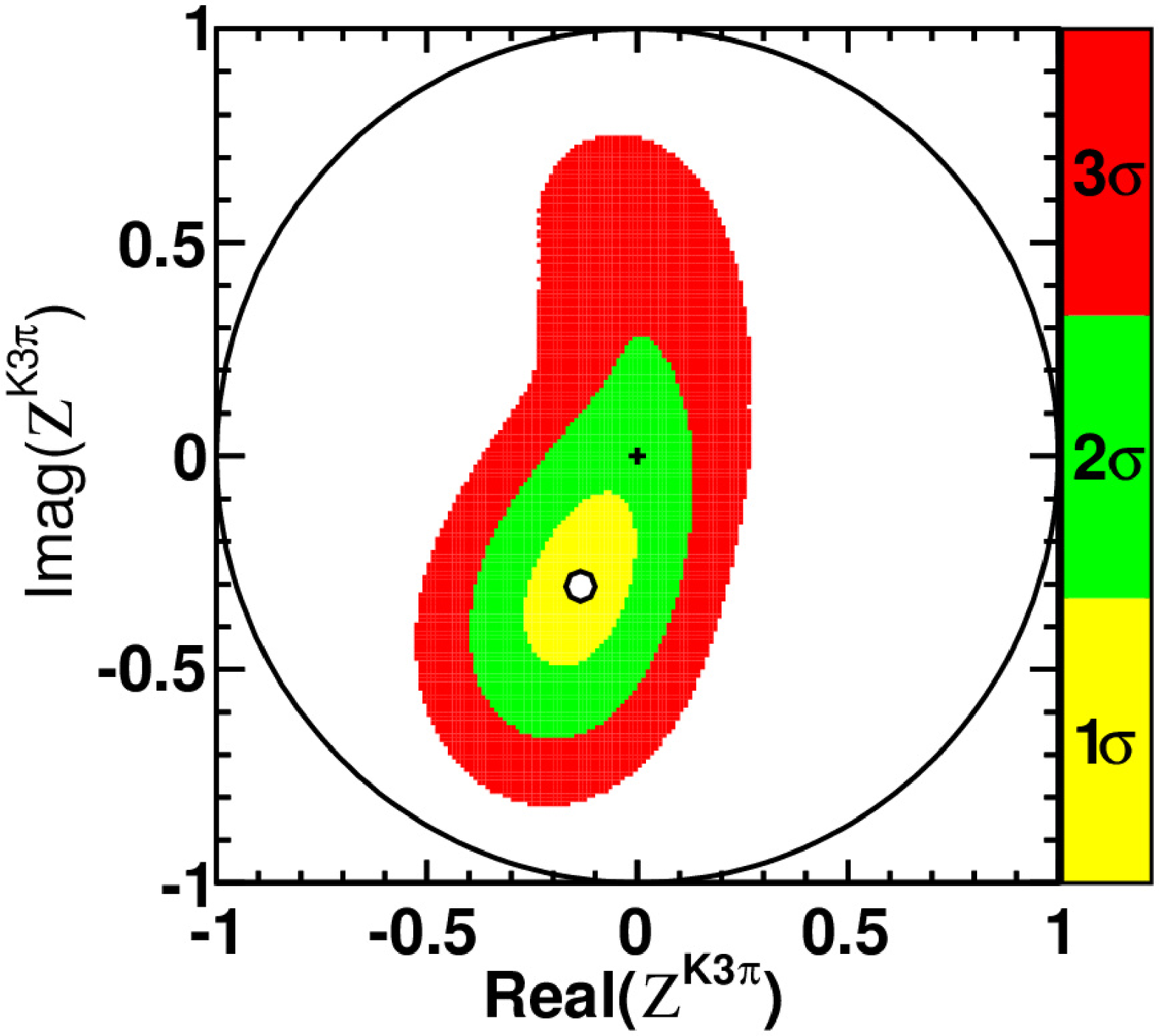}
    \end{minipage}
    \hspace{0.5cm}
    \begin{minipage}[b]{0.48\linewidth}
        \centering
        \includegraphics[width=\linewidth]{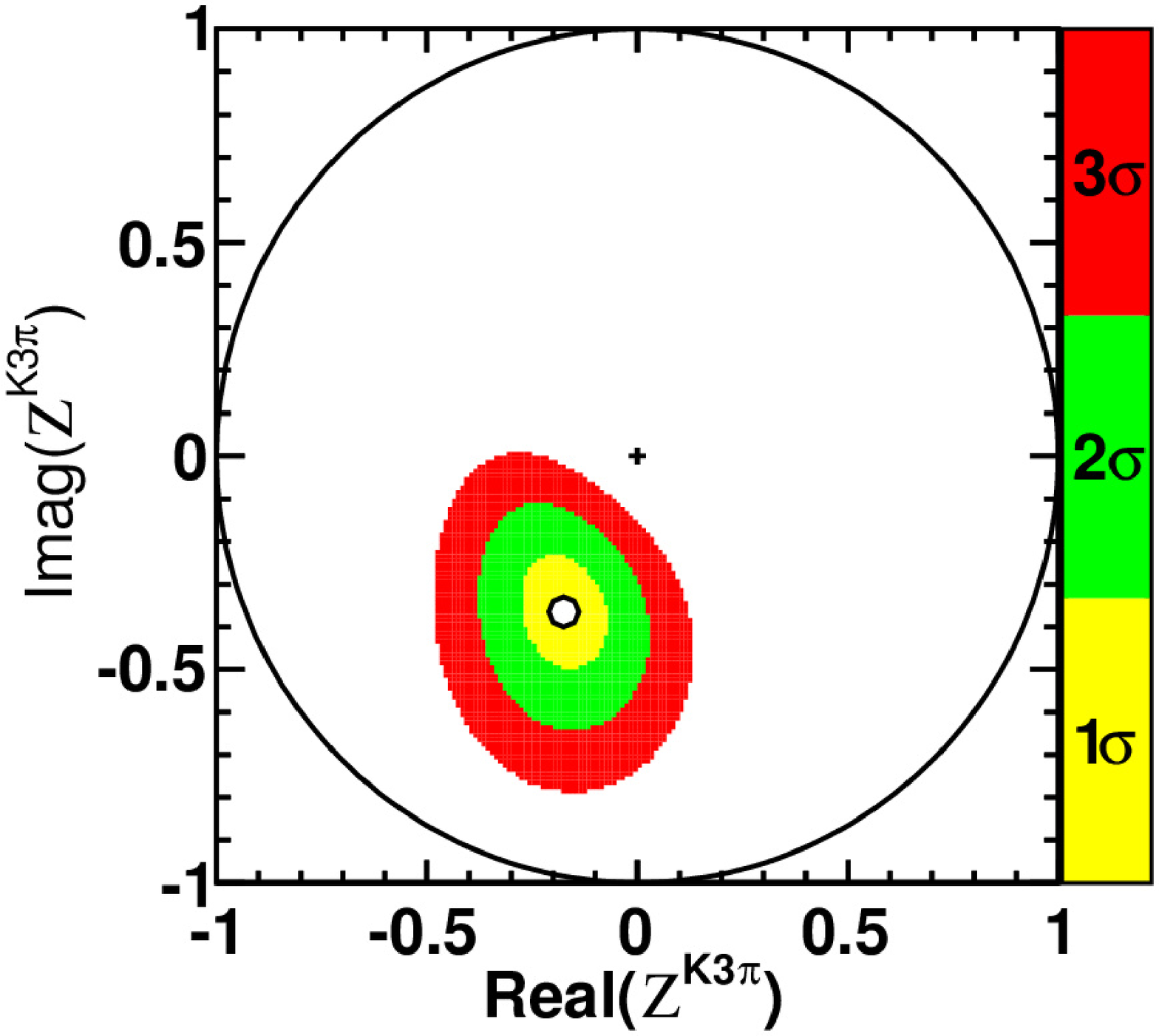}
    \end{minipage}
    \caption{Constraints on \ZKpipipi\ obtained by
      \cleoc~\cite{Lowery:2009id} are shown
      on the left. Constraints obtained by combining the \cleoc\
      results with the input
      from simulated \DtoKpipipi\ charm mixing data are shown on the right. The
      simulated signal sample is
      similar in size to that expected from \un{3}{fb^{-1}} of data taken by
      LHCb in 2011 and 2012. The same results are shown in polar
      coordinates \RKpipipi,
      \delKpipipi\ (top row) and in cartesian coordinates \ReZKpipipi,
      \ImZKpipipi\ (bottom row). The white filled circle indicates
      the location with the smallest \chisq.
      \label{fig:combination}
    }
\end{figure} 
To evaluate the potential impact of input from charm mixing on the
precision of \ZKpipipi, we combine the \chisq\ function used to obtain
\figref{fig:LHCbconstraints} with \cleoc's measurement of
\ZKpipipi~\cite{Lowery:2009id}. The \cleoc\ results, and the
combination with our simulated data, are shown in
\figref{fig:combination}. The input from charm mixing improves the
constraints considerably. The effect is particularly striking at the
$\ge 2\sigma$ level where there were previously no constraints on
\AveStrongPhaseDiff.
\begin{table}
\centering
\begin{tabular}{ c || c || c  || c }
\multicolumn{4}{c}{Fit result (where available) with $68\%$
 confidence intervals ($\Delta\chisq$)} \\
& Simulation $8M$ evts   
& \multicolumn{1}{c||}{\cleoc~\cite{Lowery:2009id}} 
& \multicolumn{1}{c}{Combination}
\\\hline
\parbox{0pt}{\mbox{}\vspace{1ex}\\\mbox{}}
$\RKpipipi$  &  $[0.28, 1.00]$     & $0.33_{- 0.23}^{+0.20}$  &  $\mbox{}\;\;\;0.40_{- 0.11}^{+0.13}$ \\
\parbox{0pt}{\mbox{}\vspace{0.8ex}\\\mbox{}}
$\delKpipipi$  &  $[1.07, 3.77]$  & $1.99_{- 0.42}^{+0.46}$ & $\mbox{}\;\;\;2.03_{- 0.27}^{+0.33}$ \\
\parbox{0pt}{\mbox{}\vspace{0.8ex}\\\mbox{}}
$\ReZKpipipi$  &  --  & $-0.14_{- 0.14}^{+0.14}$ & $-0.18_{- 0.10}^{+0.11}$ \\
\parbox{0pt}{\mbox{}\vspace{0.8ex}\\\mbox{}}
$\ImZKpipipi$  &  --  & $-0.31_{- 0.19}^{+0.23}$ & $-0.37_{- 0.14}^{+0.14}$
\\
\multicolumn{4}{c}{}
\\
\multicolumn{4}{c}{Bayesian $95\%$ confidence intervals}
\\\hline
$\RKpipipi$       &  $\mbox{}\;\;[0.27, 1.00]$  & $\mbox{}\;\;[0.00,0.63]$ & $\mbox{}\;\;[0.20, 0.66]$  \\
$\delKpipipi$  &  $\mbox{}\;\;[1.07, 3.83]$   &  --     & $\mbox{}\;\;[1.51, 2.77]$ \\
$\ReZKpipipi$  &  $[-0.96, 0.50]$  & $[-0.41,0.13]$ & $\mbox{}[-0.39,0.03]$  \\
$\ImZKpipipi$  &  $[-0.58, 1.00]$   & $[-0.69,0.41]$  & $\mbox{}\;\;\;[-0.65,-0.11]$
\end{tabular}
\caption{Constraints on \RKpipipi\ and \delKpipipi\ as
  well as \ReZKpipipi\ and \ImZKpipipi\ from
  simulation, \cleoc~\cite{Lowery:2009id},~and their combination, at
  $68\%$ and $95\%$ CL, obtained with two different techniques
  following~\cite{Lowery:2009id}, as described in the text. The $\Delta
  \chisq$ method is not suitable for obtaining separate constraints on \ReZKpipipi\ and
  \ImZKpipipi\ from the simulated mixing data alone.
 \label{tbl:constraintSummary}}
\end{table}
To quantify these improvements, one-dimensional 68\%\ and 95\%\
confidence intervals for \RKpipipi\ and \delKpipipi\ are calculated,
following the same procedures as used by \cleoc~\cite{Lowery:2009id}
to ensure comparable results. The 68\% confidence limits are based on
a standard \chisq\ difference calculation. The same process would lead
to 95\%\ confidence limits reaching the edge of the
\CoherenceFactor-\AveStrongPhaseDiff\ parameter space in the \cleoc\
measurement. These are therefore obtained using a Bayesian approach
with a uniform prior in the physically allowed region of the parameter
of interest. The results are summarised in
\tabref{tbl:constraintSummary}. The constraints from our $8M$
simulated charm events (with $30k$ WS events) shrink the existing
uncertainties on \ReZKpipipi\ and \ImZKpipipi\ by a factor of $\sim
1.5$, and the $95\%$ CL on \ImZKpipipi\ by a factor of two. In terms
of polar coordinates, the simulated input approximately halves the
uncertainty in \RKpipipi, and significantly reduces the uncertainty on
\delKpipipi. There is currently no constraint on \delKpipipi\ at the
$2\sigma$ level, and only a one-sided upper limit for \RKpipipi. From
the combination of our simulated data with the \cleoc\ result, we
obtain $\delKpipipi\ \in [1.51,2.77]$, and $\RKpipipi \in [0.20,
0.66]$ at 95\% confidence.

\section{Conclusion}
\label{sec:conclusion}
Charm mixing is sensitive to the same same charm interference
parameters that are relevant to the measurement of \gam\ in \BDK\ and
related decay modes~\cite{Atwood:coherenceFactor,
BaBar_uses_us:2011up,ModelIndepGammaTheory,LHCb2012DalitzGamma,
Malde:2011mk,Bondar:CharmMixingCP,ChrisAndGuy2012}. So far, these have
only been accessible at the charm
threshold~\cite{Libby:2010nu,Briere:2009aa,Insler:2012pm,
Lowery:2009id,CLEO:DeltaKpi}. The increased precision with which the
charm mixing parameters $x$ and $y$ have been
measured~\cite{CDF:Mixing2008,Belle:Mixing2007,BaBar:Mixing2007,BaBar:Mixing2008,BaBar:Mixing2009,LHCb:Mixing,HFAG2013,CLEO:DeltaKpi}
opens up the possibility of constraining charm interference parameters
using charm mixing. However, previous studies indicate that for decays
to self-conjugate final states, such as \DtoKspipi\ and \DtoKsKK,
datasets much larger than those currently available are required to
significantly improve constraints on the binned \catchyName{}s
$\Zi=c_i + i s_i$ from charm mixing~\cite{ChrisAndGuy2012}. On the
other hand, in wrong-sign decay modes such as \DtoKpipipiDCS\ and
\DtoKpipiZeroDCS, the mixing-induced interference effects are
significantly enhanced compared to self-conjugate decays. This
provides greater sensitivity to the \catchyName\ \Zall, or,
equivalently, the coherence factor $\CoherenceFactor=|\Zall|$ and
average strong phase difference $\AveStrongPhaseDiff = -\arg(\Zall)$
introduced in~\cite{Atwood:coherenceFactor}. While it is interesting
to note that useful information can be obtained in this way without
additional input, the true power of the method lies in the combination
with threshold data. We evaluate the potential of this approach with a
simulation study based on estimated \DtoKpipipi\ signal yields expected
in LHCb's 2011 and 2012 dataset. We do not assume any improvements
on external inputs. Our results indicate that charm mixing input from
existing LHCb data, when combined with \cleoc's
measurement~\cite{Lowery:2009id}, could substantially reduce the
current uncertainty on the coherence factor and average strong phase
difference in \DtoKpipipi. Such a measurement can be expected to have
a significant impact on the precision with which the CKM parameter
$\gam$ can be measured at LHCb, BELLE~II, and the LHCb upgrade.

\section*{Acknowledgements}
\label{sec:acknowledgements}
\noindent We thank our colleagues at CLEO-c and LHCb for their helpful
input to this paper, in particular Tim Gershon, Jim Libby, Andrew
Powell and Guy Wilkinson. We also acknowledge support from CERN, the
Science and Technology Facilities Council (United Kingdom) and the
European Research Council under FP7.




\bibliographystyle{elsarticle-num}
\bibliography{bibliography}






\end{document}